\documentstyle[12pt]{article}
\title{\bf INERTIA AS EMERGING OF A HAMILTONIAN CONSTRAINT SYSTEM}
\author{\bf Farhad Darabi\thanks{e-mail: f-darabi@cc.sbu.ac.ir}\\
\\
{\small Department of Physics, Shahid Beheshti University, Evin, Tehran 19839, Iran.} \\
\\
\\
{\small mailing address:} \\ 
{\small Department of Physics} \\
{\small Shahid Beheshti University}\\ 
{\small Evin, Tehran 19839} 
{\small Iran.} \\
}
\begin{document}
\maketitle
\vspace{15mm}
\begin{abstract}
The issue of inertia as opposition to acceleration of a massive point particle 
in Minkowski space-time is investigated in the context of a Hamiltonian constraint 
system. It is shown that the inertia as a locally-originating force in Minkowski 
space-time may emerge due to a global constraint.
\end{abstract}
\newpage
\section{Introduction}

Recently, the issue of inertia as an ``unresolved mystery in modern physics'' 
has been the subject of intense investigations. Disagreement of Mach's principle 
with general relativity is the well known fact that have been presented,  
with additional conflicts, by Vigier and Rindler in recent papers \cite{VR}.  
In his paper, Vigier has used the Dirac vacuum to explain, in an alternative 
way, the origin of inertia. More recently, Rueda $et\:al.$ in a series of papers 
\cite{R} have explained the inertia as scattering-like process 
of the ZPF radiation subject to the electromagnetic vacuum. 
 
In this paper, we shall adopt a different approach to explain the possible origin 
of inertia based on a Hamiltonian constraint structure. 
To this aim, we concentrate on a massive point particle as an elementary constituent 
of matterial objects for which we shall investigate the issue of inertia. In this 
way, inertia of a matterial object may be described as well. The motivation in 
following this approach is the belief that the inertia is a locally-originating force 
in Minkowski space-time, emerging of an inherent global (independent of space-time ) 
constraint in a causal structure.
In our opinion, inertia as an inherent property of matter do not depend on the  
collective linkage of all matter in the universe according to Mach's idea, instead 
it is just a local  effect (even in an empty space-time) which emerge identically 
all over the space-time due to the existence of an inherent global constraint. 
As is well known, the action of a relativistic free massive point particle has a 
Hamiltonian constraint structure \cite{Dirac} leading to the mass-shell condition, 
where the space-time  coordinates are assumed to be dynamical variables subject 
to the Minkowski metric. 
Of course, this constraint structure is independent of the space-time dimensions,  
so one may be motivated to reconstruct an equivalent 
constraint structure without the need to use the space-time coordinates as dynamical 
variables. We use the worldline of a point particle as 
dynamical variable and parametrize its action by a prefered time parameter, 
then we construct the associated constraint structure which is   
independent of space-time dimensions as a global one.\footnote{This is because 
the worldline can be defined in arbitrary dimensional space-time with any arbitrary 
metric.}. 
It turns out that inertia as opposition to acceleration of the point particle 
in Minkowski space-time may emerge as a consequence of this constraint structure.

\vspace{5mm}

\section{Constraint structure}

We start with the action of a relativistic point particle 
\begin{equation}
I = - m_0 c \int\!d{\bf s} 
\end{equation}
where $m_0$ is the rest mass of the particle, $c$ is the velocity of light 
and $d{\bf s}$ the worldline element. 
We now rewrite the action as
\begin{equation}
I = - m_0 c \int_{\theta_1}^{\theta_2}\!\dot{{\bf s}} d\theta   
\end{equation}
where $\dot{{\bf s}} = \frac{d{\bf s}}{d\theta}$ and $\theta$ is assumed to be a prefered 
time parameter. The question that who observes the velocity $\dot{{\bf s}}$ will be 
answered later. 
This linear action admits a Hamiltonian constraint structure \cite{Dirac} with the primary constraint
\begin{equation}
\phi \equiv p_s + m_0 c\approx 0
\end{equation}
which shows the reparametrization invariance of the action (2), where 
$p_s$ is the momentum conjugate to ${\bf s}$.
The original Hamiltonian $H_0$ is zero by reparametrization invariance, so the
total Hamiltonian is
\begin{equation}
H_T = \lambda(\theta) \: \phi
\end{equation}
where $\lambda(\theta)$ is an arbitrary function of time; hence the dynamics is fully 
controlled by the constraint (3). The consistency condition
\begin{equation}
0 \approx \dot{\phi} = \{\phi,H_T\} 
\end{equation}
is satisfied since $\{\phi,\phi\} \approx 0$, so that there is 
no secondary constraints. The global
\footnote{The constraint (3) is global in the sense that it is independent of  
local space-time coordinates.} constraint (3) is by definition \cite{Dirac} a first class 
constraint, which generates the following infinitesimal gauge transformations
\begin{equation}
\delta {\bf s} = \epsilon(\theta) \{{\bf s},\phi\} = \epsilon(\theta)
\end{equation}
\begin{equation}
\delta p_s = \epsilon(\theta)\{p_s,\phi\} = 0
\end{equation}
where $\epsilon(\theta)$ is an infinitesimal arbitrary function of time.
Gauge invariance of the action (2) holds as long as $\epsilon(\theta_1)=\epsilon(\theta_2)$. 
It follows from (6) and (7) that the variable ${\bf s}$ has a gauge orbit and so is not a 
measurable quantity, where the momentum $p_s$ is a global gauge invariant 
quantity on this orbit and is measurable. Considering the constraint (3) reveals 
the fact that the whole physical content of this theory, after gauge fixing, is 
based on this gauge invariant degree of freedom in phase space since as will be 
seen, it carries an important characteristic of a point particle namely its compton wavelength. 
\newline
The equations of motion are
\begin{equation}
\dot{{\bf s}} = \{{\bf s},H_T\} = \lambda(\theta)
\end{equation}
\begin{equation}
\dot{p_s} = \{p_s,H_T\} = 0
\end{equation}
which integrate to
\begin{equation}
{\bf s} = \int^{\theta}\!\lambda(\theta')\:d{\theta'}
\end{equation}
\begin{equation}
p_s = {\mbox Const.}
\end{equation}
these are consistent with (6) and (7) provided that $\delta \lambda = \dot{\epsilon}(\theta)$.

Quantization of this system is done by operating on the Hilbert subspace $\mid\psi>$ 
by the constraint operator (3) as \cite{Dirac}
\begin{equation}
\hat{\phi}\:\mid\psi> = 0
\end{equation}
which gives a time independent wave function
\begin{equation}
\psi({\bf s}) \sim e^{-i\frac{m_0 c}{\hbar} {\bf s}}
\end{equation}
It is remarkably interesting to note that the compton wavelength of the point 
particle $\frac{\hbar}{m_0 c}$, is the natural length scale measuring its worldline length.

\section{Equivalence classes}

Once the previous constraint structure is constructed, one sees that all arbitrary 
motions of a massive point particle in Minkowski space-time correspond to arbitrary  
dynamics $\dot{{\bf s}}=\lambda(\theta)$ subject to the constraint $\dot{p_s}=0$. Therefore 
the types of functions $\lambda(\theta)$ should be defined properly. 
By looking at the action (1) it turns out that the only natural velocity in the 
theory is the constant velocity of light $c$. We assume that the  
observer, who shall observe the velocity $\dot{{\bf s}}$, does not 
recognize whether a massive point particle is at rest or in a uniform motion in 
Minkowski space-time. This assumption is based on the equivalence 
of all inertial motions in Minkowski space-time. 
Therefore this observer have to associate a unique constant velocity $\bar{\lambda}=c$ 
for all massive point particles at rest or in uniform motion in Minkowski space-time. 
It has its origin in 
the fact that the existence of another constant velocity $\bar{\lambda} \neq c$ 
will give rise to appearence of an ambiguity in the compton wavelength, because 
there would not be any physical way to recognize $a\:priori$ which of 
$\frac{\hbar}{m_0 c}$ or $\frac{\hbar}{m_0 \bar{\lambda}}$ is the compton wavelength,  
so, this breaks down the constraint (3)
\footnote{It appears that once we fix the factor $m_0 c$ in the action, then the   
unique constant velocity $c$ is prefered by which every massive point particle 
at rest or in uniform 
motion in Minkowski space-time is forced to move, as wived by this observer.}. 
According to above discussion we 
define two types of functions $\lambda(\theta)$ as follows\\
1) Constant function $\bar{\lambda}=c$ corresponds to all uniform motions of a 
massive point particle in Minkowski space-time.\\
2) Each time dependent function $\lambda(\theta)$ corresponds to each non-uniform 
motion of a massive point particle in Minkowski space-time.\\
Two classifications are based on the fact that one can distinguish physically 
between a uniform motion and non-uniform motion by resorting to inertial effects 
such as ``inertia'' of point particle and that we expect a correspondence between 
$\lambda(\theta)$ and appearence of inertia. Using $\delta\lambda=\dot{\epsilon}(\theta)$ 
and $\delta {\bf s}=\epsilon(\theta)$ we can correspondingly define two types of gauge 
equivalent motions of a massive point particle in Minkowski space-time as follows \\
1) $\delta\bar{\lambda}=0$ corresponds to a constant function $\bar{\epsilon}$ 
and therefore a constant gauge transformation $\delta {\bf s}=\bar{\epsilon}$.\\
2) $\delta\lambda(\theta)= \bar{\lambda}-\lambda(\theta)\neq 0$ corresponds to time 
dependent function $\epsilon(\theta)$ and therefore a time dependent gauge 
transformation $\delta {\bf s}=\epsilon(\theta)$.\\
The first expresses that the equivalence relation between two arbitrary different 
uniform motions of a same point particle in Minkowski space-time can be represented 
as the constant gauge transformation $\delta {\bf s}=\bar{\epsilon}$ with arbitrary 
constant $\bar{\epsilon}$.
On the other hand, one finds from the second one an equivalence relation between a 
uniform motion $\bar{\lambda}=c$ and an arbitrary non-uniform motion $\lambda(\theta)$ 
of a same point particle in Minkowski space-time which can be represented as the 
time dependent gauge transformation $\delta {\bf s}=\epsilon(\theta)$ with arbitrary 
$\epsilon(\theta)$.
However, there is a big difference between two gauge equivalences (transformations). 
The first gives rise to a gauge invariant action where the second one does not, since in 
the later case there is no requirment to have $\epsilon(\theta_1)=\epsilon(\theta_2)$. 
Therefore, the second type gauge transformations does not minimizes the action. 
This property of the action as different responses against different gauge transformations 
represents remarkably the well known fact that the action becomes minimum only for uniform motions.

\section{A prefered frame}

Comming back to the constraint (3) we concentrate on its physical content. 
As is well known, the compton wavelength of a point particle with rest mass $m_0$ 
is $\frac{\hbar}{m_0 c}$ which is fixed whenever one fixes the rest mass of that 
particle. Indeed this character is independent of any motion of the particle in 
space-time. On the contrary, we know de Broglie wavelength $\frac{\hbar}{p}$ 
depending on the momentun $p$ of the point particle in space-time. Comparing 
the compton and de Broglie wavelengthes, a natural question arises:
``Is there any kind of intrinsic momentum associated with the rest 
point particle, giving its compton wavelength ? or, Can we find a 
prefered frame in which a rest point particle in Minkowski space-time 
appears to have a momentum whose de Broglie wavelength gives compton 
wavelength of that particle ?'' 

Obviously no rest mass $m_0$ can move with the light velocity $c$ or have a 
momentum $m_0 c$ in Minkowski space-time. But, considering the constraint (3) 
turns out that this rest mass have this momentum.  
Then a remarkable result is obtained:
``One can define a prefered observer who observes this intrinsic momentum of point particle 
in his prefered frame and finds its compton wavelength as de Broglie wavelength''.

From the wievpoint of this prefered observer every massive point particle,   
at rest or in uniform motion in Minkowski space-time, is forced to move with  
the unique constant velocity $c$ and $\dot{{\bf s}}=\lambda(\theta)$ is its instantaneous 
velocity upon acceleration in Minkowski space-time. So, in order to have 
a gauge invariant (constant) momentum $m_0 c$, the change $c\rightarrow \dot{{\bf s}}(\theta)$ 
must be compensated by the corresponding change $m_0 \rightarrow m\equiv m_0 \frac{c}{\dot{{\bf s}}(\theta)}$. 
We believe that the concept of inertia as the instantaneous opposition to acceleration 
of a massive point particle can be described through this procedure occuring in this  
prefered frame.

\section{Inertia}

Here we are going to investigate the issue of inertia in the context of our 
constraint structure.  
In our opinion the concept known as inertia is based on the effects occuring from 
the wievpoint of our prefered observer, since the issue of inertia is highly 
correleted with the acceleration and makes an absolute line of demarcation between 
the uniform motion and acceleration; it is natural to investigate the inertia 
in a prefered frame in which there is also a similar demarcation. 
The equation of motion (9) combined with the constraint (3) gives rise to
\begin{equation}
\frac{d}{d\theta}(m_0 c)=0
\end{equation}
which is trivial for the uniform motion giving rise to no dynamical effect. 
However, for non-uniform motion it becomes 
\begin{equation}
\frac{d}{d\theta}(p_s)=m\frac{d\dot{\bf s}}{d\theta}+\frac{d m}{d\theta}\dot{\bf s}=0
\end{equation}
or
\begin{equation}
m\frac{d\dot{\bf s}}{d\theta}=-\frac{d m}{d\theta}\dot{\bf s}
\end{equation}
Dynamically, we note that the dimensions of both sides are the dimension of force, 
so it is a statement of Newton's third law about a symmetry in nature as
\footnote{Note that $\dot{{\bf s}}$ decreases with $\theta$ where $m$ increases with 
it, so the directions of $\vec{{\bf f}}_{ap}$ and $\vec{{\bf f}}_{re}$ are opposite.}
\begin{equation}
\vec{{\bf f}}_{ap}=-\vec{{\bf f}}_{re}
\end{equation}
with 
\begin{equation}
\vec{{\bf f}}_{ap}\equiv m\frac{\vec{d \dot{\bf s}}}{d\theta}\:\:\:  ,\:\:\:\vec{{\bf f}}_{re}\equiv \vec{\dot{\bf s}} \frac{d m}{d\theta}
\end{equation}
Therefore, this symmetry finds its truly origin in the global constraint (3). To accelerate 
a point particle the force $\vec{{\bf f}}_{ap}$ must be applied by an agent and the agent 
will necessarily experience an equall and opposite force $\vec{{\bf f}}_{re}$ as long as 
the acceleration continues. Therefore the reaction force $\vec{{\bf f}}_{re}$, corresponds to inertia 
of the point particle. It is seen that the origin of inertia in this wievpoint is the constraint dynamics, 
in particular a dynamical mass $ \frac{d m}{d \theta}$.
\\
\\
\\
{\large {\bf Discussion}} \vskip8pt\noindent

A careful insight turns out that our prefered observer have to be one with 
zero gauge transformation (translation) $\delta {\bf s}=0$. In other words,this observer has  
a zero velocity $\dot{{\bf s}}=0$ or $d{\bf s}=0$ which is the  
main characteristic of massless particles. Hence, it seems that the  
concept of inertia is interrelated with the existence of massless particles 
moving with light velocity in Minkowski space-time such as photons.  
In this way, the relativistic action (2) by which the global constraint (3) is derived, preserves 
the causality as the underlying basis of Newton's third law and the concept of inertia.  
In this respect, inertia is a locally-originating force in Minkowski space-time  
due to the existence of the global constraint (3). We believe that the approach considered here, may have  
an implicit relation with that of works explaining the origin of inertia by 
an acceleration-dependent scattering of ZPF momentum flux \cite{R}. This is because in both approaches the 
existence of massless particles moving with the velocity of light plays a key role 
explaining inertia in a way that preserves the causality. Moreover, the electromagnetic 
drag effect with the steady growth of the ZPF momentum contained within the 
volume of the object, in that works, may be identified with dynamical mass $\frac{d m}{d\theta}$ 
obtained here as the origin of inertia. On the other hand, the existence of a
prefered frame associated with the massless particles, as a requirement to describe
the inertia in this work, may remind us to look at the relation to Dirac's ether
model implying inertial forces assuming the existence of the absolute local inertial
frames associated with the observed local isotropy of 2.7 K background microwave
radiation \cite{D}.


\newpage
\begin{thebibliography}{99}
\bibitem{VR}J. P. Vigier, Foundations of Physics, {\bf 25}, No. 10, 1461 (1995); \\
W. Rindler, Phys. Lett. A {\bf 187}, 236 (1994).
\bibitem{R}A. Rueda and B. Haisch, $ Inertia\:as\:reaction\:of\:the\:vacuum\:
to\:accelerated\:motion$\\ Phys. Lett. A, in press;\\
A. Rueda and B. Haisch, Foundations of Physics, in press (1998);\\
B. Haisch, A. Rueda and H. E. Puthoff, Phys. Rev. A {\bf 49}, 678 (1994).
\bibitem{Dirac}P. A. M. Dirac, $Lectures \:on\: Quantum \:Mechanics$, 
Yeshiva University (Academic-Press, New York 1967).  
\bibitem{D}P. A. M. Dirac, Nature (London) {\bf 168}, 906 (1951);\\
M. C. Combourieu and J. P. Vigier, Phys. Lett. A {\bf 175}, 269 (1993).
\end{thebibliography}
\end{document}